\documentclass[11pt,twoside]{article}

\usepackage{epsfig}

\usepackage{amsmath}
\usepackage{amssymb}
\usepackage{cite}

\begin{document}
\newcommand{\pst}{\hspace*{1.5em}}

\newcommand{\rigmark}{\em Journal of Russian Laser Research}
\newcommand{\lemark}{\em Volume ***, Number ***, 201***}

\newcommand{\be}{\begin{equation}}
\newcommand{\ee}{\end{equation}}
\newcommand{\bm}{\boldmath}
\newcommand{\ds}{\displaystyle}
\newcommand{\bea}{\begin{eqnarray}}
\newcommand{\eea}{\end{eqnarray}}
\newcommand{\ba}{\begin{array}}
\newcommand{\ea}{\end{array}}
\newcommand{\arcsinh}{\mathop{\rm arcsinh}\nolimits}
\newcommand{\arctanh}{\mathop{\rm arctanh}\nolimits}
\newcommand{\bc}{\begin{center}}
\newcommand{\ec}{\end{center}}

\thispagestyle{plain}

\label{sh}


\begin{center} {\Large \bf
TRANSMISSION OF CORRELATED GAUSSIAN PACKETS THROUGH A DELTA-POTENTIAL
 } \end{center}

\bigskip

\bigskip

\begin{center} {\bf
V. V. Dodonov and A. V. Dodonov 
}\end{center}

\medskip

\begin{center}
{\it
Instituto de F\'isica, Universidade de Bras\'ilia, 70919-970  Bras\'ilia,  DF, Brasil
}
\smallskip

E-mails:~~~vdodonov@fis.unb.br\\
\end{center}

\begin{abstract}\noindent
We study the evolution of the most general initial Gaussian packet 
with nonzero correlation coefficient between the coordinate and momentum operators
in the presence of a repulsive delta potential barrier, using the known exact propagator of the
time-dependent  Schr\"odinger equation. For the initial packet localized far enough from the barrier,
we define the transmission coefficient as  the probability of discovering the particle in the whole semi-axis
on the other side of the barrier. It appears that the asymptotical transmission coefficient (calculated
in the large time limit) depends on two dimensionless parameters: the normalized ratio of the potential strength
to the initial mean value of momentum and the ratio of the initial momentum dispersion to the initial
mean value of momentum. For small values of the second parameter the result is reduced to the well
known formula for the transparency of the delta barrier, obtained in the plane wave approximation
by solving the stationary Schr\"odinger equation. For big values of the second parameter, the transmission
coefficient can be much bigger than that calculated in the plane wave approximation. 
For a fixed initial spread of the packet in the coordinate space, the initial correlation coefficient 
influences the transparency of the barrier only indirectly, through the
increase of the initial momentum dispersion.
 \end{abstract}

\medskip


\section{Introduction}
\pst

Among many amusing properties of the quantum world, perhaps, the most well known and exciting examples are 
the {\em uncertainty relations\/} \cite{Heis27,Kennard} and the {\em tunneling effect} \cite{Gamow,Gurney29}.
In  the simplest forms they can be expressed by the known formulas 
\be
\Delta x \Delta p \ge \hbar/2
\label{Heis}
\ee
 and
\be
D \sim \exp\left(-2\int_1^2\sqrt{2m\left[U(x) -E\right]}dx/\hbar\right),
\label{Gamow}
\ee
 where $\hbar$ is the Planck constant, $m$ is the mass of particle,
$D$ is the transmission probability through the potential barrier $U(x)$, and the meaning of other symbols 
is well known.  

But is is not so well known for many people (although not for all, of course) that (\ref{Heis}) is a special case of a more general inequality,
discovered independently by Schr\"odinger \cite{Schr} and Robertson \cite{Rob},
\be
\sigma_x \sigma_p -\sigma_{xp}^2 \ge \hbar^2/4,
\label{SR}
\ee
where 
\be
\sigma_x =\langle\left(x-\langle x\rangle\right)^2\rangle ,  \qquad
\sigma_p =\langle\left(p-\langle p\rangle\right)^2\rangle , \qquad
\sigma_{xp}= \langle \hat{x}\hat{p} + \hat{p}\hat{x}\rangle/2 - \langle x\rangle\langle p\rangle.
\ee
It was noticed in \cite{DKM79} that inequality (\ref{SR}) can be rewritten in the form
\be
\sigma_x \sigma_p \ge \frac{\hbar^2}{4\left(1-r^2\right)},
\label{UR-r}
\ee
where
\be
r =  \frac{\sigma_{xp}}{\sqrt{\sigma_{x}\sigma_{p}}}
\label{r}
\ee
is the {\em correlation coefficient\/} between the coordinate and momentum.
Looking at the right-hand side of inequality (\ref{UR-r}), one is tempted to interpret it as $\hbar_{ef}^2/4$,
with the {\em effective Planck constant\/}
\be
\hbar_{ef}= \frac{\hbar}{\sqrt{1-r^2}}.
\label{hef}
\ee
Then a natural question arises: Does such an interpretation has any physical meaning, at least in some
specific situations, or is it a mere speculation? Or in other words: Are there situations (quantum systems) where the nonzero correlation
coefficient can lead to real physical effects? One of the most impressive consequences could be  \cite{DKM79,Chernega13} a significant
increase of the tunneling probability, if the true Planck constant $\hbar$ could be replaced in formula (\ref{Gamow}) 
by the effective constant (\ref{hef}) or by some other effective constant.
Some attempts to answer these questions were made in 
\cite{DOM-192,DKM-200,DKM-PLA96,Campos99}, and recently this problem was studied in the series of papers
\cite{Vysot10,Vysot12a,Vysot12b,Vysot13}.

However, it seems that the full answers have not been found until now. 
Here we try to give a partial answer, considering, perhaps, the simplest possible example: the propagation of the initial Gaussian
correlated state through a delta-potential barrier (in one dimension). The reason for this choice and its advantage consist in the
possibility to solve the time-dependent Schr\"odinger equation exactly, so that only a few integrals giving the final result need numerical
calculations. 

We dedicate this paper to the memory of Allan Solomon.
We both met him for the first time in 1990 in Moscow during the XVIII International Colloquium on
Group Theoretical Methods in Physics, where one of the authors spoke about the physical meaning of correlated states \cite{DKM-90G}.
Since then we had a pleasure to meet Allan at many scientific and private events in different countries and continents, enjoying his
brilliant talks and stories. We believe that this article would be interesting for him.

\section{Evolution of the correlated Gaussian packet}
\pst

We consider a general Gaussian form of the initial free wave packet,
\begin{equation}
\psi (x,0)=(\pi s^{2})^{-1/4}\exp \left[
-\frac{(x-x_{c})^{2}\gamma}{2s^{2}} - ip_0 x/\hbar \right] , \qquad
\gamma = 1-i\rho, 
 \qquad x_{c} \gg s.
\label{psiindel}
\end{equation}
This packet has the initial mean value of the coordinate $\langle x\rangle_0=x_c >0$.
 The initial mean value of the momentum is supposed to be negative,
$\langle p\rangle_0 = -p_0$ (with $p_0>0$), so that the packet moves in the negative direction of the $x$ axis.
The initial  variances of coordinate and momentum are given by the formulas
\be
\sigma_x(0) =\langle\left(x-x_c\right)^2\rangle = s^2/2, \qquad
\sigma_p(0)= \langle\left(p-p_0\right)^2\rangle = \hbar^2\left(1+\rho^2\right)/\left(2s^2\right), 
\ee
and their covariance equals
\be
\sigma_{xp}(0)= \langle \hat{x}\hat{p} + \hat{p}\hat{x}\rangle/2 - x_c p_0=\hbar\rho/2,
\ee
so that  the Schr\"odinger--Robertson uncertainty relation (\ref{SR}) becomes an equality \cite{Kennard}. 
The 
correlation coefficient (\ref{r}) is related to the parameter $\rho$ as follows,
\be
r  = \frac{\rho}{\sqrt{1+\rho^2}}, \qquad 
\rho = \frac{r}{\sqrt{1-r^2}}, \qquad -\infty <\rho < \infty, \quad -1 < r <1.
\ee
The state  (\ref{psiindel}) was named {\em correlated coherent state\/} in  \cite{DKM79}). 
Later, similar packets describing the motion of a free particle were considered
under the name ``contractive states'' in \cite{Yuen83,Stor94}.
The dynamics of Gaussian wavepackets possessing nonzero correlation
coefficients was studied in \cite{Hell91}.
Gaussian correlated packets were used in the analysis of many physical problems,
from the wave propagation in optical waveguides \cite{Kriv86} and random media \cite{Sin89} to
the neutron interferometry \cite{Ler95}, cosmology \cite{Barv95} and quantum optics
\cite{Campos99,Walls96}.
Detailed studies of properties and dynamics of correlated and squeezed coherent states in arbitrary time-dependent
harmonic potentials were performed in
\cite{DOM-192,DKM-200,176,183,dodonov2,Sukhanov02}.
Recently some special time dependent harmonic oscillator potentials were considered in
\cite{Vysot10,Vysot12a,Vysot12b,Vysot13}.

Here, our aim is to calculate the evolution of the initial state  (\ref{psiindel})  in the presence of the potential
$V(x)={\cal Z}\delta (x)$. Explicit formulas describing the propagation of Gaussian packets through the delta potential
barrier were obtained earlier by several authors  \cite{Janos52,Naka97,AD04,Villav07,Cord10}, but for zero initial correlation coefficient only.
The evolution of any initial wave function $\psi(x,0)$ is determined by the propagator
$G(x,x^{\prime };t)$ according to the relation
\begin{equation}
\psi (x,t)=\int_{-\infty }^{\infty }G(x,x^{\prime };t)\psi (x^{\prime},0)dx^{\prime }.
\label{psi0-psit}
\end{equation}
The integral representation for the propagator in the presence of delta-potential was found in
\cite{Gav,Law88,Manouk89}:
\be
G_{\cal Z}(x,x^{\prime };t) = G_{0}(x,x^{\prime };t)
- {\cal Z}\left( 2\pi i t\right)^{-1/2} \int_0^{\infty} du
\exp \left[ -u {\cal Z} -  \left(|x|+|x^{\prime }|+u\right)^2/(2it) \right],
\label{prop-int}
\ee
where
\be
G_{0}(x,x^{\prime };t) =
\left( 2\pi i t\right)^{-1/2}
\exp \left[ i(x-x^{\prime })^{2}/(2 t)\right]
\label{G0}
\ee
is the well known free particle propagator.
To simplify formulas, we assume hereafter that $m=\hbar=1$. 
The return to true dimensional variables can be performed by means of the replacements
\be
t\to {\hbar t}/{m}, \qquad p \to p/\hbar, \qquad {\cal Z} \to {m{\cal Z}}/{\hbar^2}.
\quad
\label{dim}
\ee
The integral in the right-hand side of (\ref{prop-int}) can be expressed in terms of the complementary
error function \cite{Bauch,Blind,Elber88}, but for our purposes the integral representation seems more
convenient. Putting expressions (\ref{psiindel}), (\ref{prop-int}) and (\ref{G0}) in formula (\ref{psi0-psit}),
we calculate first the integral over $x^{\prime }$, and after that we calculate the integral over the auxiliary
variable $u$.
Taking into account that the initial function $\psi (x^{\prime},0)$ is exponentially small 
in the region $x^{\prime}<0$ under the condition $x_{c} \gg s$, we replace $|x^{\prime}|$ by $x^{\prime}$ 
in the integrand of (\ref{prop-int}) even for $x^{\prime}<0$. Then the integral over $x^{\prime}$ 
becomes Gaussian in whole axis $-\infty < x^{\prime } < \infty $, so it can be calculated exactly. 
After that, the integration over $u$ can be performed with the aid of formula
\be
\int_0^{\infty} \exp\left(-ax^2 +bx\right)dx=
\frac12\sqrt{\frac{\pi}{a}}\exp\left(\frac{b^2}{4a}\right){\rm erfc}\left(\frac{-b}{2\sqrt{a}}\right),
\label{int-erfc}
\ee
which holds for complex parameters $a$ and $b$, provided ${\rm Re}(a)>0$ \cite{Grad}.
The complementary error function is defined as
\begin{equation}
{\rm erfc}(z)=\frac{2}{\sqrt{\pi }}\int_{z}^{\infty }\exp \left(-y^{2}\right) \,dy\equiv 1-{\rm erf}(z)  .
\label{def-erfc}
\end{equation}%
The final result can be represented as follows,
\be
\psi(x,t)=\psi_{free}(x,t)\left\{ 1 - {\cal Z}\sqrt{\frac{\pi\mu(t)}{2\gamma}}{\rm erfc}[D(x,t)]
\exp\left[D^2(x,t) +\frac{x+|x|}{\mu(t)}\left(ip_0 s^2 -\gamma x_c\right)\right]
\right\},
\label{psit-f}
\ee
where
\be
\mu(t) = s^2 + i\gamma t, \qquad
D(x,t) =\left[{\cal Z}\mu(t) +\gamma\left( |x| + x_c\right)  -ip_0 s^2 \right]/\sqrt{2\gamma \mu(t)},
\label{def-D}
\ee
and
\be
\psi_{free}(x,t) = \left[\sqrt{\pi}\mu(t)/s\right]^{-1/2}\exp\left[ -\frac{\gamma}{2\mu(t)}
\left(x-x_c +p_0 t\right)^2 -ip_0 x -ip_0^2 t/2 \right]
\label{psifree}
\ee
is the wave function of freely moving Gaussian packet in the absence of external potentials
(remember that the initial mean value of momentum was $-p_0$).

\section{Transmission coefficient}
\pst

The crucial point for the following analysis is the observation that ${\rm Re}(D) \gg 1$ under the conditions $x_c \gg s$
and ${\cal Z} >0$ (a repulsive barrier). This is obvious for $t=0$, when $\mu=s^2$. 
Fot $t \to\infty$, this is also true, due to the leading term ${\cal Z}\sqrt{\mu}$ in the formula for $D(x,t)$, 
whose real part grows as ${\cal Z}\sqrt{t}$ in this limit.
Therefore, we can use the asymptotical formula \cite{Grad}
\begin{equation}
\mbox{erfc}(x)\approx \frac{\exp (-x^{2})}{x\pi ^{1/2}}, \qquad
|x|\rightarrow \infty,  \qquad |\arg x|<\frac{3\pi }{4},
\label{erfcass}
\end{equation}
which permits us to simplify formula (\ref{psit-f}) for $x<0$ (when $x+|x| \equiv 0$) as follows:
\be
\psi(x<0,t) \approx \psi_{free}(x,t) \frac{\gamma\left( |x| + x_c\right)  -ip_0 s^2}
{\gamma\left( |x| + x_c\right)  -ip_0 s^2 +\mu(t){\cal Z}}.
\label{psi-simple}
\ee
Now we notice that function $|\psi_{free}(x,t)|^2$ is concentrated in the vicinity of its maximum at the
point $x_m=x_c - p_0 t$. Then, introducing the new variable $y$ according to the formula
$x=x_c + p_0 t(y-1)$, we find the following asymptotical coordinate probability density for $t \to \infty$
(and for a fixed value of $x_c$): 
\be
{\cal P}(y)  \equiv \left|\psi(x_c + p_0 t(y-1);t\to\infty)\right|^2 = \left|\psi_{free}^{as}(y,t)\right|^2 
\frac{(1-y)^2}{(1-y)^2 +A},
\label{Py}
\ee
where
\be
\left|\psi_{free}^{as}(y,t)\right|^2 \equiv \left|\psi_{free}(x_c + p_0 t(y-1);t\to\infty)\right|^2 
= \left[t\sqrt{\pi\left(1+\rho^2\right)}/s\right]^{-1} \exp\left[-\,\frac{(sp_0 y)^2}{1+\rho^2}\right]
\label{psifas}
\ee
and
\be
A = \left({\cal Z}/p_0\right)^2 \equiv \left(\frac{m{\cal Z}}{\hbar p_0}\right)^2.
\label{A}
\ee
For $y=0$, we obtain the free probability density multiplied by the well known plain wave transmission coefficient 
through the delta barrier
\be
T_{A}= (1+A)^{-1}.
\label{TA}
\ee
But some deformations of the free packet shape are observed  for $y\neq 0$.
\begin{figure}[htb]
\includegraphics[width=0.5\textwidth]{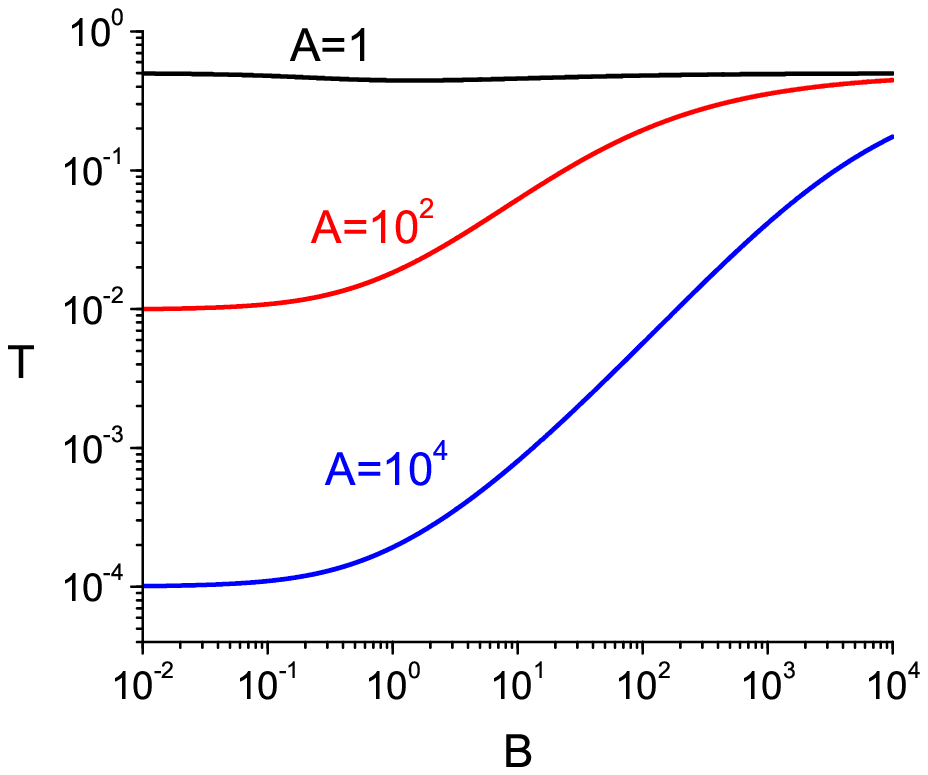} 
\hfill%
\includegraphics[width=0.5\textwidth]{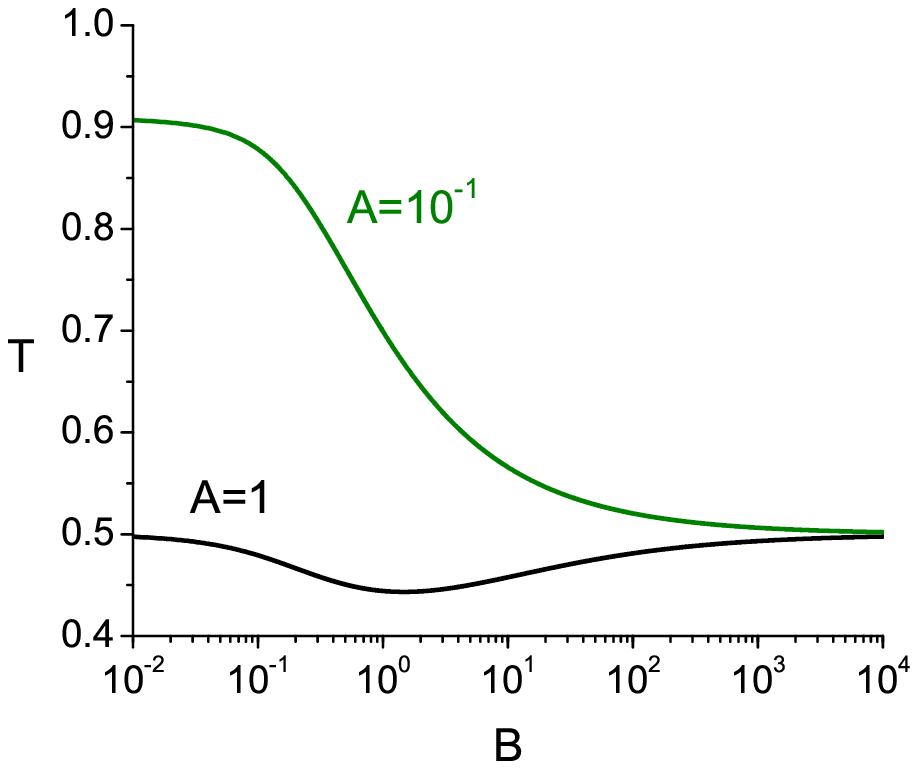}%
\vspace{-4mm}
\caption{\label{fig-TAB}  The transmission coefficient through the delta barrier as a function of parameter $B=\sigma_p(0)/p_0^2$  for
fixed values of the normalized strength of the potential $A=\left({\cal Z}/p_0\right)^2$.}
\end{figure}

Defining the asymptotical transmission coefficient as the integral of the asymptotical coordinate probability density over the semi-axis
$x<0$ (past the delta barrier), we arrive at the formula (taking into account that $dx =tp_0 dy$)
\be
T(A,B)= (2\pi B)^{-1/2} \int_{-\infty}^1 \exp\left(-\frac{y^2}{2B}\right) \frac{ (1-y)^2 dy}{(1-y)^2 +A},
\label{TAB}
\ee
where
\be
B = \frac{1+\rho^2}{2(sp_0)^2} =\frac{\sigma_p(0)}{p_0^2}.
\label{B}
\ee
We see that although the transmission coefficient depends on $\rho$ in principle, the crucial parameter is not $\rho$ itself, but the ratio
of the initial momentum dispesrsion $\sqrt{\sigma_p(0)}$ to the initial mean value of momentum $p_0$. If this ratio is much smaller than unity,
i.e., $B\ll 1$, then the integral in formula (\ref{TAB}) is determined by the small region of the order of $\sqrt{B}$ nearby the origin.
Then the pre-exponential fraction can be replaced by its value at $y=0$ (if $A$ is not extremely small), and the integration can be extended
formally from $-\infty$ to $\infty$. In this case, $T(A,B)=T(A)$. 

Another special case, when the integral in (\ref{TAB}) can be calculated
approximately analytically, is $1\ll B \ll A$. Then the term $(1-y)^2$ in the denominator of the fraction can be neglected, and the integral
is reduced to $\int_0^{\infty} y^2 \exp\left[-y^2/(2B)\right]$ (plus small corrections). In this case, 
$T(A,B) \approx B/(2A) =\left( 1+\rho^2\right)/(2s{\cal Z})^2$. 
Although this
expression is proportional to $1+\rho^2=\left(1-r^2\right)^{-1}$, nonetheless we cannot say that it can be obtained from the plane wave
transmission coefficient by the replacement $\hbar \to \hbar_{ef}$, since in the latter case we would obtain from formulas
(\ref{A}) and (\ref{TA}) the new formula $T_{ef}=\left( 1+\rho^2\right)\left(p_0/{\cal Z}\right)^2$.

In Fig. \ref{fig-TAB}, we show the dependence of the transmission coefficient on the parameter $B$ for different fixed values of
the parameter $A$.
We see that the plane wave transmitivity formula can be used with a good accuracy if $B\le 0.1$. 
For bigger values of $B$, the transmission coefficient grows monotonously if $A>1$, 
but it decreases as a function of $B$ if $A<1$. 
$T(A,B)$ is practically constant for $A=1$, with a small decrease (about $10\,$\%) for $B=1$.
Asymptotically,
$T(A,B) \to 1/2$ when $B\to \infty$. This is explained simply by the fact that the initial wave packet is so wide in the momentum space 
for $B\gg 1$, that the actual probabilty to move in the direction of the barrier is only $1/2$.

\begin{figure}[htb]
\includegraphics[width=0.6\textwidth]{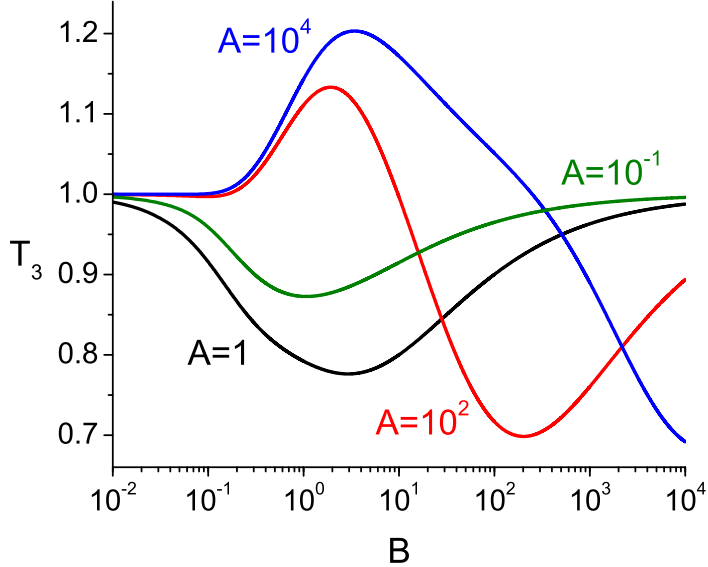} 
\vspace{-4mm}
\caption{\label{fig-apr}  The ratio $T_3 \equiv T(A,B)/T_{apr}(A,B)$ of the exact transmission coefficient 
(\ref{TAB}) to the interpolating formula (\ref{Tapr})
 as a function of parameter $B=\sigma_p(0)/p_0^2$  for
fixed values of the normalized strength of the potential $A=\left({\cal Z}/p_0\right)^2$.}
\end{figure}

Trying to guess an analytical interpolation for the function $T(A,B)$, we can suppose that it can be obtained by replacing 
 $p_0^2$ in the plain wave formula (\ref{TA}) with the square of the total initial momentum $\langle p^2 \rangle = p_0^2 + \sigma_p(0)$.
In addition, we have to take into account the non-unity probability of moving in the direction of the barrier, which equals
${\rm erfc}\left(-{1}/{\sqrt{2B}}\right)/2$.  In this way, we obtain the formula
\be
T_{apr}(A,B)= \frac12\left(1 + \frac{A}{1+B} \right)^{-1} {\rm erfc}\left(-\frac{1}{\sqrt{2B}}\right)
=\frac1{\sqrt{2\pi B}} \int_{-\infty}^1 \exp\left(-\frac{y^2}{2B}\right) dy \left(1 + \frac{A}{1+B} \right) ^{-1}.
\label{Tapr}
\ee
Its comparison with the exact values of $T(A,B)$ is given in Fig. \ref{fig-apr}.
We see that the interpolation formula (\ref{Tapr}) is reasonable qualitatively, although it can under-estimate or over-estimate
the correct transmission coefficient up to $30\,$\%.
This means that, strictly speaking, the transmission coefficient does not depend on the sum $p_0^2 + \sigma_p(0)$,
but it depends on $p_0^2$ and $\sigma_p(0)$ separately.

\section{Discussion}
\pst

We have seen that the transmission coefficient of the most general correlated Gaussian wave packet through a delta potential
barrier depends (for a fixed strength of the potential) on the ratio of the initial momentum dispersion to the initial mean
momentum, but it does not depend on the correlation coefficient alone. In particular, formulas for the transmission
coefficient, derived for the incident plane waves, are valid under the condition $\sigma_p \ll p_0^2$, even if the correlation
coefficient is relatively big.
Therefore, it seems that the idea about an ``effective Planck constant''
does not work in the case of tunneling of free packets. This can be understood from the observation that the effective
Planck constant appears in formula (\ref{UR-r}), which contains the variances of the coordinate and momentum only. 
Since the first order mean values, such as $\langle x \rangle$ and  $\langle p \rangle$, are independent from the variances,
one should not expect the appearance of the effective Planck constant in formulas containing the mean value $\langle p \rangle$,
 in particular, in formulas for the tunneling probabilities of free packets impinging potential barriers.
Another reason, why the correlation coefficient alone cannot influence the tunneling probability is that the correlation
coefficient in any quantum state of a free particle depends on time, as well as the coordinate dispersion. Therefore,
physical results cannot depend on the initial values of these two quantities separately. In contrast, the momentum variance
is preserved during the free propagation, so it can influence physical results, together with the momentum mean value.
However, we cannot exclude a possibility that the effective Planck constant can enter the formulas for the escape probabilities
from some potential wells with two barriers, since in such cases the mean value $\langle p \rangle$ equals zero
for many initial states localized between the barriers (in particular, in the cases studied in Refs. 
\cite{DKM-PLA96,Vysot10,Vysot12a,Vysot12b,Vysot13}).  But we leave a detailed analysis of such cases for another study.

\section*{Acknowledgments}
\pst
The authors acknowledge a partial support of the Brazilian agency CNPq. 
We thank Prof. V. I. Man'ko for discussions that stimulated this research.

\end{document}